\documentstyle[amsfonts,epsfig,12pt,cite]{article}
\newcommand{\vb}[1]{{\mathbf{#1}}}

\newcommand{\lb}[1]{\label{#1}}

\newcommand{\bc}{\begin{center}}
\newcommand{\ec}{\end{center}}
\newcommand{\be}{\begin{equation}}
\newcommand{\ee}{\end{equation}}
\newcommand{\bea}{\begin{eqnarray}}
\newcommand{\eea}{\end{eqnarray}}
\newcommand{\ba}[1]{\begin{array}{#1}}
\newcommand{\ea}{\end{array}}

\newcommand{\bt}[1]{\begin{table}[ht]\centering\begin{tabular}{#1}}
\newcommand{\et}[1]{\end{tabular}\caption{\small#1}\end{table}}

\newcommand{\diag}{{\mathrm{diag}}\,}

\begin{document}


\thispagestyle{empty}

\begin{center}

\vspace{1.3 truecm}

{\large\bf{Mass for Plasma Photons from Gauge Symmetry Breaking}}
\vspace{1.4 truecm}\\

{\bf J. T. Mendon\c{c}a}\\[2mm]
{\small{\it CFP and CFIF, Instituto Superior T\'ecnico, 1096 Lisboa, Portugal\\[5mm]
Centre for Fundamental Physics, Rutherford Appleton Laboratory, Chilton Didcot, Oxfordshire, UK}}\\[5mm]
{\bf P. Castelo Ferreira}\\[2mm]
{\small{\it CENTRA, Instituto Superior T\'ecnico, Av. Rovisco Pais, 1049-001 Lisboa, Portugal\\[5mm]
Departamento de F\'{\i}sica, Universidade da Beira Interior, Rua Marqu\^es D'\'{A}vila e Bolama, 6200-081 Covilh\~a, Portugal}}\\[1.5cm]

{\bf\sc Abstract}\\[5mm]

\begin{minipage}{16cm}
We derive the effective masses for photons in unmagnetized plasma waves using
a quantum field theory with two vector fields (gauge fields). In order to properly define the
quantum field degrees of freedom we re-derive the classical wave equations on light-front gauge.
This is needed because the usual scalar potential of electromagnetism is, in quantum field theory,
not a physical degree of freedom that renders negative energy eigenstates. We also consider
a background local fluid metric that allows for a covariant treatment of the problem.
The different masses for the longitudinal (plasmon) and transverse photons are in our
framework due to the local fluid metric. We apply the mechanism of mass generation
by gauge symmetry breaking recently proposed by the authors by giving a non-trivial
vacuum-expectation-value to the second vector field (gauge field). The Debye length
$\lambda_D$ is interpreted as an effective compactification length and we compute an
explicit solution for the large gauge transformations that correspond to the specific
mass eigenvalues derived here. Using an usual quantum field theory canonical quantization we
obtain the usual results in the literature. Although none of these ingredients are new to
physicist, as far as the authors are aware it is the first time that such constructions
are applied to Plasma Physics. Also we give a physical
interpretation (and realization) for the second vector field in terms of
the plasma background in terms of known physical phenomena.\\[5mm]

Addendum: It is given a short proof that equation (10) is wrong, therefore equations (12-17) are
meaningless. The remaining results are correct being generic derivations for nonmagnetized
plasmas derived in a covariant QFT framework.

\end{minipage}

\end{center}

\vfill
\begin{flushleft}
\end{flushleft}

\newpage

\setcounter{page}{1}

\section{Introduction}

It is well known that in Plasma physics the photons acquire an effective mass. For unmagnetized plasmas,
this mass is associated with a lower frequency cut-off that is determined by the plasma frequency.
The main question raised is if this effect can be described by some fundamental underlying theory~\cite{Tito_01}.
This question is important, not only for a deeper understanding of Plasma physics itself, but also for a more
general description of massive vector fields. Here we address the photon mass issue from
a point of view of quantum field theory given by a covariant variational principle.

There are two main issues addressed in this paper. Firstly, we re-derive the usual plasma wave
equations in order to obtain the usual Proca equations~\cite{Proca}. In order to achieve it
we are going to use a non Coulomb gauge, this is necessary because the usual scalar potential $\phi=A_0$
is not a physical degree of freedom in quantum field theory and upon quantization it renders negative
energy eigenstates. Specifically we use a light-front gauge $A_+=0$~\cite{lc,lcqed}.
Also in order to express the equations in a covariant form we consider a longitudinal
plus transverse spatial decomposition and introduce a background local fluid metric that allows to
deal with the theory in a covariant way. This construction have been originally applied to Bose-Einstein
condensates by Unruh~\cite{Unruh}.
Secondly, we use the mechanism proposed by the authors in~\cite{Proca_mass} in order to
generate a mass to the photons from a variational principle. Here we explicitly compute the
large gauge transformations for our specific problem and obtain an explicit diagonal mass matrix.
Finally we canonically quantize the theory to show that indeed our quantum field theory approach does
hold the same results of the semi-classical approach to plasma waves that have been previously
considered in the literature (see~\cite{Tito_01} and references therein).

\section{The Massive Wave Equations for the Photon\lb{sec.wave}}

Usually in Plasma Physics the Coulomb gauge $\nabla.A=0$ is used. In this gauge the space components
of the gauge fields (the usual vector potential $A_i$) are therefore divergentless and the time component
of the gauge fields (the usual scalar potential $A_0$) is a total divergence ($\nabla\times A_0=0$) such that
in the absence of magnetic fields the divergence of the
electric field is given by $\nabla.E=-\nabla.(\nabla A_0+\partial_0 A)=-\nabla^2A_0$.
Therefore the longitudinal waves or plasmons are expressed uniquely in terms of $A_0$.
At classical level this does not raises a problem. However at quantum field theory level
the $A_0$ is not a physical degree of freedom, even worst, it gives a negative contribution to
the Hamiltonean. This is due to our space-time being Minkowski and $A_0$ being a time-like
degree of freedom, such that upon canonical quantization
the commutators are proportional to the metric $[a_\mu^\dag,a_\nu]\sim g_{\mu\nu}$. For massless photons
this problem is solved by considering the Coulomb gauge with the choice $A_0=0$ (such that $A_0$ plays
the role of a Lagrange multiplier that imposes the Gauss law) or the Lorentz gauge $\partial_\mu A^\mu$
in which case the $A_0$ and $A_\parallel$ (the longitudinal component) degrees of freedom combine together
being excluded from the physical spectrum (see~\cite{ryder} for a detailed discussion) and only the
transverse degrees of freedom survive. For massive photons however gauge symmetry is broken 
and we have three massive degrees of freedom, two transverse and one longitudinal~\cite{Schwinger,Anderson}.
Then using Coulomb gauge clearly raises a problem concerning the usual description of
longitudinal waves in terms of $A_0$ and the respective quantization
since the condition $\nabla.A=0$ only allows for
two space-like degrees of freedom. Here we will solve this
problem by working in light-front gauge $A_+=0$~\cite{lc,lcqed} and considering a decomposition
into longitudinal and transverse spatial directions. Also in order to cast the
usual classical equations in a covariant form we introduce a background local fluid metric~\cite{Unruh}.
We note that generally a fluid-front gauge is enough and holds the same results.

We consider a space decomposition into longitudinal and transverse directions such that
\be
\ba{rcccl}
z&=&\displaystyle\frac{c^2}{S_e^2}x^3&=&\mathrm{longitudinal\ direction}\\[5mm]
x_\perp&=&x^i&=&\mathrm{transverse\ directions}
\ea
\ee
where $z$ is the longitudinal rest-frame coordinate, $x^3$ the laboratory longitudinal coordinate
and $i=1,2$ stand for the two transverse directions.
For the derivatives we use the convention $\partial_\parallel=\partial_z$
for the longitudinal derivative and $\nabla_\perp=(\partial_1,\partial_2)$ for
the transverse gradient operator.

As explained in detail in the introduction, in order to consistently quantize the theory we use the gauge
\be
A_+=\frac{1}{c}A_z+A_t=0\ \ \ \ \Leftrightarrow\ \ \ \  A_z=-cA_t\ .
\ee
Then the longitudinal electric field is
\be
E=F_{tz}=(c\partial_t+\partial_z)A_z
\ee

As shown in~\cite{Tito_01} we obtain from the Maxwell equations
\be
\ba{rcl}
\displaystyle\partial_z(c\partial_t+\partial_z)A_z&=&\displaystyle\frac{e}{\epsilon_0}n\\[5mm]
\displaystyle\left(\nabla_\perp^2-\frac{1}{c^2}\partial_t^2\right)A_\perp&=&\displaystyle J_\perp
\ea
\lb{Maxwell}
\ee
with $J_\perp=-(e^2/n_0)A_\perp$ and the equation of motion for the plasma electrons
\be
\left(\partial_t^2-S_e^2\partial_z^2\right)n=\frac{e\,n_0}{m}\partial_z(c\partial_t+\partial_z)A_z
\lb{electrons}
\ee
the equations for the longitudinal field $A_z$ and the two transverse fields $A_\perp$ read
\be
\ba{rcl}
\displaystyle\left(\partial_z^2-\frac{1}{S_e^2}\partial_t^2\right)A_z&=&\displaystyle\frac{\omega_p}{S_e^2}A_z\\[5mm]
\displaystyle\left(\nabla^2_\perp-\frac{1}{c^2}\partial_t^2\right)A_\perp&=&\displaystyle\frac{\omega_p}{c^2}A_\perp\ .
\ea
\lb{waves}
\ee
The first equation can be obtained by considering a Fourier transformation of the first Maxwell
equation~(\ref{Maxwell}) and~(\ref{electrons}) and solving for $A_z$~\cite{plasmas}. $S_e=3v_e$,
being $v_e=\sqrt{T/m}$ the thermal velocity of electrons.

Also we note that as explained in~\cite{Tito_01} the transverse modes and the longitudinal modes
do not have the same masses which is manifest in the energy spectrum upon canonical quantification.
Furthermore we are working in the photon rest frame and that so far we have not specified
what is the explicit form of the metric we are working with. By considering the above equations~(\ref{waves})
with a rescaling of the longitudinal coordinate $z=(S_e/c)x^3$ we can cast the equations in a covariant form
holding the usual Proca equation for the spatial components of the gauge field (vector potential)
\be
\partial_\mu\partial^\mu A_i=\frac{\omega_p}{c^2}A_i\ \ \ \ \ \ i=1,2,3\ .
\lb{Proca_eq}
\ee

Therefore the local fluid metric corresponding to equations~(\ref{waves}) is
\be
ds^2=-c^2(dt)^2+(dx^1)^2+(dx^2)^2+\frac{c^2}{S_e^2}(dz)^2=-(dx^0)^2+(dx^1)^2+(dx^2)^2+(dx^3)^2\ .
\lb{metric}
\ee
For simplicity from now on we will work in the coordinate system $(x^0,x^1,x^2,x^3)$ corresponding to the
metric $\eta={\mathrm{diag}}(-1,1,1,1)$. We note that this metric is local in the sense that only describes
geometric properties of the space-time, as we have shown in~\cite{Proca_mass} the topology of the space is
relevant in our construction, in particular it will be relevant that for plasma waves there exist an effective
compactification length along the longitudinal direction that is given by the Debye length $\lambda_D$.

\section{Effective Electric Action and the Mass Matrix for Longitudinal Vacua\lb{sec.mass}}

Now we will describe the mass matrices from an effective action point of view using
the mass generation mechanism proposed by the authors in~\cite{Proca_mass}.
We use the effective electric theory of~\cite{Proca_mass}
\be
S^{(e)}_{Eff}=\int dx^4\left[-\frac{1}{4e^2}F_{\mu\nu}F^{\mu\nu}+\frac{1}{2}A_\mu\,M^{\mu\nu}\,A_\nu\right]
\lb{Seff_Electric}
\ee
with the mass matrix given by
\be
M^{\mu\nu}=-\frac{2}{e^2}\left[\partial^\mu\tilde{c}^\nu+\frac{1}{2}\left(g^{\mu\nu}\,\tilde{c}^\alpha\tilde{c}_\alpha-\tilde{c}^\mu\tilde{c}^\nu\right)\right]
\ee
which is obtained by the breaking of a $U(1)\times U(1)$ gauge symmetry trough a vacuum expectation values
for the field $\tilde{c}$ and respective derivatives.

The respective equations of motion varying the above action are
\be
\partial_\alpha\partial^\alpha A^\mu=-M^{\mu\nu}A_\nu\ \ \ \ \ \ \mu=0,1,2,3\ .
\lb{eom}
\ee
The mass matrix $M_{\mu\nu}$ is not diagonal, however we can generally bring it
to a diagonal form by a suitable coordinate transformation. We consider only longitudinal
dependence on the vacua such that the transverse and time elements of the mass
matrix vanish, i.e. $\tilde{c}^0=\tilde{c}^1=\tilde{c}^2$.

Following~\cite{Proca_mass} we have for some large gauge transformation $V=\exp\{if(z)\}$
\be
\ba{rcl}
\tilde{c}^z&=&2f'(z)\\[5mm]
\partial^z\tilde{c}^z&=&2f''(z)\ .
\ea
\ee
In this particular case the mass matrix is diagonal
\be
M^{\mu\nu}=\diag\left(\frac{(\tilde{c}^z)^2}{e^2},-\frac{(\tilde{c}^z)^2}{e^2},-\frac{(\tilde{c}^z)^2}{e^2},-\frac{2}{e^2}\partial^z \tilde{c}^z\right)\ .
\lb{mass_matrice}
\ee
Solving the differential equation
\be
f''(z)=2(f'(z))^2
\ee
we obtain
\be
f(z)=f_0-\frac{1}{2}\ln\left\{2z+\alpha\right\}
\ee
where $f_0$ and $\alpha$ are integration constants. This solution hold the expectation values
\be
\left<f''(z)\right>=2\left<(f'(z))^2\right>=\frac{1}{2\alpha}-\frac{1}{2(\alpha+2\lambda_D)}\ .
\lb{f_values}
\ee
Here $\lambda_D$ stands for the Debye wavelength $\lambda_D=v_e\,\omega_p$ and the expectation values
are computed as the integral between $0$ and $\lambda_D$. We can interpret
this length as a natural effective compactification of the $z$ coordinate such that effectively
we have a filled torus with only one holonomy cycle (the $z$ coordinate).

Solving~(\ref{f_values}) for $\alpha$ in order to obtain the desired mass eigenvalues of~(\ref{Proca_eq})
we obtain two possible values
\be
\alpha_\pm=-v_e\left(\omega_p\pm\sqrt{1+\frac{c^2}{v_e e^2}}\right)\ .
\ee
Then we have
\be
A_\mu M^{\mu\nu}A_\nu=+\frac{\omega_p}{c^2}A_0^2-\frac{\omega_p}{c^2}\left(A_1^2+A_2^2+A_3^2\right)
\ee
which correctly have a minus sign for the spatial component and a plus sign for the time component
in order that the equations of motion~(\ref{eom}) to give a positive mass.

\section{Canonical Quantization in Rest-Frame\lb{sec.quant}}
Canonical quantization proceeds now in the standard way. Using the Lorentz condition $\partial_\mu A^\mu=0$ and using
it to eliminate $A_0$ from the action we obtain the canonical conjugate moments
\be
\pi^0=0\ \ \ ,\ \ \ \pi^i=-\partial_0A^i\ .
\ee
Imposing the usual equal-time commutation relations
$[A^i(\vb{x}),\pi_j(\vb{x}')]=i\hbar\delta^i_{\ j}\delta(\vb{x}-\vb{x}')$ we obtain
\be
[A_i(\vb{x}),\dot{A}_j(\vb{x}')]=i\hbar\,g_{ij}\delta(\vb{x}-\vb{x}')\ \ \ \ \ i,j=1,2,3\ .
\ee
Expanding the fields in Fourier modes as usual
\be
A_{\mu,k}=\int\frac{d^3\vb{k}}{(2\pi)^3\sqrt{2k_0}}\sum_{\lambda=1}^3\varepsilon_{\mu,\vb{k}}^{\lambda}\left[a_{\lambda,\vb{k}}e^{-i\vb{k}.\vb{x}}+a_{\lambda,\vb{k}}^\dag e^{+i\vb{k}.\vb{x}}\right]
\ee
we obtain the commutation relations 
\be
[a_{\lambda,\vb{k}}^\dag,a_{\lambda',\vb{k}'}]=g_{\lambda\lambda'}\hbar \sqrt{2k_0}(2\pi)^3\delta(\vb{k}-\vb{k}')\ .
\ee
The polarization vectors are space like obeying as usual the conditions $\varepsilon^\lambda.\varepsilon^{\lambda'}=g^{\lambda\lambda'}$
and $\varepsilon^\lambda.k=0$ such that we have in the rest-frame
\be
\ba{rcl}
k^\mu&=&\displaystyle\left(\frac{\omega_p}{c},0,0,0\right)\\[5mm]
\varepsilon^1&=&\displaystyle\left(0,1,0,0\right)\\[5mm]
\varepsilon^2&=&\displaystyle\left(0,0,1,0\right)\\[5mm]
\varepsilon^3&=&\displaystyle\left(0,0,0,\frac{c}{S_e}\right)
\ea
\ee
where we used the metric for the rest-frame as given in~(\ref{metric}).
Therefore after normal ordering we obtain the same Hamiltonian of~\cite{Tito_01}
\be
H=\hbar\int\frac{d^3\vb{k}}{(2\pi)^3}\sum_{\lambda=1}^3\omega_{\lambda,\vb{k}}\ a_{\lambda,\vb{k}}^\dag a_{\lambda,\vb{k}}
\ee
where we subtracted the ground state energy and the frequency $\omega_{\lambda,k}=\sqrt{w_p+k^2c_\lambda^2}$
with $c_1=c_2=c$ and $c_3=S_e$ are obtained from the usual mass-shell condition.

\section{Conclusions}

In this work we have addressed the problem of the origin of an effective mass for photons and plasmons in
unmagnetized plasmas using a quantum field theory with two vector fields (gauge fields).
In order to properly define the quantum field degrees of freedom we have re-derived the classical
wave equations on light-front gauge and considered an underlying local fluid metric. We note however
that a fluid-front gauge would render the same results. The different masses for the longitudinal
(plasmons) and transverse photons are in our framework due to this local fluid metric.
We have applied successfully the mechanism of mass generation by gauge symmetry breaking
recently proposed by the authors in~\cite{Proca_mass}. By considering an effective
compactification length given by the Debye length $\lambda_D$ we manage to compute
explicit solutions for the large gauge transformations that render non-trivial
vacuum-expectation-values for the second gauge field. In this way we have obtained an explicit
diagonal mass matrix. Carrying a usual quantum field theory canonical quantization we obtain the
same results of other works in the literature that use a semi-classical approach~\cite{Tito_01}.

Although none of the mechanisms used here are new, as far as we are aware it is the first time that they
are applied to Plasma Physics. It is also interesting to note that the action we are considering is
compatible with the existence of pure physical magnetic charges~\cite{Dirac,CF} and that as shown
in~\cite{action_00,PCF}, by consistence imply the existence of two physical vector fields (gauge fields).
In here we manage, for the first time, to give a physical interpretation to the second vector field,
i.e. the plasma background.

{\bf Acknowledgements}\\
Work of PCF supported by SFRH/BPD/17683/2004.

\newpage
\nonumber

\begin{center}
{\bf Addendum to Mass for Plasma Photons from Gauge Symmetry Breaking \, [Europhys. Lett. {\bf 75} (2006) 189-194]\,}\\[4mm]

{P. Castelo Ferreira\\ \small pedro.castelo.ferreira@ist.utl.pt\\ \small CENTRA, Instituto Superior T\'ecnico, Av. Rovisco Pais 1, 1049-001 Lisboa, Portugal}\\[4mm]
\end{center}

The manuscript~\cite{epl} considers the effects of quartic terms
in the gauge theory $U_e(1)\times U_g(1)$ that generate a plasmon mass matrix in
nonmagnetized plasmas. Although the calculations and derivations in this work
are correct the main new result is based in an incorrect analisys in the unpublished
e-print~\cite{eprint}.

The specific mass matrix obtained in~\cite{eprint} and given in equation~(10) of~\cite{epl}
is not derivable within $U_e(1)\times U_g(1)$. The derivation of this expression requires
the existence of quartic terms (on the gauge fields) which, as claimed in~\cite{eprint},
is due to considering the gauge covariant derivative
\be
D_\mu=\partial_\mu-A_\mu-\hat{\epsilon}\tilde{C}_\mu\ .
\lb{D}
\ee
The gauge invariant field tensor (also known as gauge connection or gauge
curvature) corresponding to this covariant derivative is
\be
{\mathcal{F}}_{\mu\nu}=\left[D_\mu,D_\nu\right]=D_\mu\,D_\nu-D_\nu\,D_\mu=-F_{\mu\nu}-\hat{\epsilon}\tilde{G}_{\mu\nu}\ ,
\ee
where we are considering a real representation for the Lie algebra of the fields. Then, considering
the definition $\tilde{G}^{\mu\nu}=\epsilon^{\mu\nu\lambda\rho}G_{\lambda\rho}/2$
and the identity $\epsilon^{\alpha\beta\mu\nu}\epsilon_{\alpha\beta\lambda\rho}=-2(\delta^\mu_{\ \lambda}\delta^\nu_{\ \rho}-\delta^\mu_{\ \rho}\delta^\nu_{\ \lambda})$, we obtain the Lagrangean density
\be
\ba{rcl}
{\mathcal{L}}&=&\displaystyle-\frac{1}{4}{\mathcal{F}}_{\mu\nu}{\mathcal{F}}^{\mu\nu}\\[5mm]
&=&\displaystyle-\frac{1}{4}F_{\mu\nu}F^{\mu\nu}-\frac{1}{4}\tilde{G}_{\mu\nu}\tilde{G}^{\mu\nu}-\frac{\hat{\epsilon}}{2}F_{\mu\nu}\tilde{G}^{\mu\nu}\\[5mm]
&=&\displaystyle-\frac{1}{4}F_{\mu\nu}F^{\mu\nu}+\frac{1}{4}G_{\mu\nu}G^{\mu\nu}-\frac{\hat{\epsilon}}{2}\epsilon^{\mu\nu\lambda\rho}F_{\mu\nu}G_{\lambda\rho}\ ,
\ea
\ee
which is the usual Lagrangean for $U_e(1)\times U_g(1)$ theory~\cite{math} and we conclude that
no quartic terms are present. Hence no mass matrix is present in the Lagrangean and the
equations~(12) to~(17) in~\cite{epl} are meaningless due to being derived based in equation~(10)
of the manuscript. As for the remaining equations are still valid corresponding to generic results derived for nonmagnetized
plasmas using a covariant approach within the framework of quantum field theory and matching the
ones derived in~\cite{quant}.

Work supported by SFRH/BPD/34566/2007.

\end{document}